\documentstyle[12pt,aaspp4]{article}

\begin{document}

\title{Dark Matter and Gas Concentrations in the NGC$\,$4532 / DDO$\,$137 System}

\author{G. Lyle Hoffman}
\affil{Dept. of Physics, Lafayette College, Easton, PA  18042; hoffmang@nova.phys.lafayette.edu}

\author{Nanyao Y. Lu}
\affil{IPAC, Mail Stop 100-22, California Institute of Technology, Pasadena, CA 91125; lu@ipac.caltech.edu}

\author{E.E. Salpeter}
\affil{Center for Radiophysics and Space Research, Cornell University, Ithaca, NY  14853; salpeter@spacenet.tn.cornell.edu}

\and

\author{Bryan M. Connell}
\affil{Dept. of Physics, Lafayette College, Easton, PA  18042; connellb@lafayette.edu}

\slugcomment{To appear in {\em AJ} February 1999}

\begin{abstract}

\ion{H}{1} synthesis mapping of NGC$\,$4532 and DDO$\,$137, a pair of Sm galaxies on the edge of the Virgo cluster,  is used to determine rotation curves for each of the galaxies and to resolve the structure and kinematics of three distinct \ion{H}{1} concentrations embedded in an extended envelope of diffuse \ion{H}{1} discovered in earlier Arecibo studies of the system.
The \ion{H}{1} masses of the three concentrations do not appear to be sufficient for them to be self-gravitating; however, their \ion{H}{1} masses and dynamical masses are very similar to those of faint Im galaxies in the Virgo cluster.
The peak \ion{H}{1} column density of each of the clouds is close to the star formation threshold, but CCD images in B and R reveal no trace of stars nor star formation.
If the system is gravitationally bound and in Virial equilibrium, we find its total mass to be ten times that within the outermost \ion{H}{1} contours of the individual galaxies and clouds.
Thirty-seven percent of the total \ion{H}{1} lies outside the boundaries of the individual galaxies as determined by the VLA mapping; 28\% cannot be ascribed to either galaxy nor to the discrete concentrations.
One possible way to account for the velocity field and the large quantity of diffuse gas is to assume that DDO$\,$137, the gas concentrations and other apparent tidal features are due to gas infall into clumps within the dark matter potential of the group, and that prior to the interaction the group consisted of an irregular galaxy (NGC$\,$4532) and a large, star-poor \ion{H}{1} cloud like \ion{H}{1}$\,$1225+01 (Giovanelli et al. \markcite{GWH91}1991).

\end{abstract}


\section{Introduction}

Studies of galaxy pairs and small groups in many cases reveal extended bridges, streams and detached clouds of \ion{H}{1} at considerable distances from the stellar components of the interacting galaxies (e.g., Haynes et al. \markcite{HGR79}1979; Haynes \markcite{Ha81}1981; van Moorsel \markcite{vM88}1988; van Driel et al. \markcite{vD+92}1992; Chengalur et al. \markcite{CST94}1994; Li \& Seaquist \markcite{LS94}1994; Yun et al. \markcite{YHL94}1994; Elmegreen et al. \markcite{EKBES95}1995b; Hibbard \& van Gorkom \markcite{HvG96}1996; Kaufman et al. \markcite{KBETESK97}1997; Nordgren et al. \markcite{NCST97a}1997a,\markcite{NCST97b}b; earlier references are cited in each of those papers).
The streams and clouds are generally assumed to be generated tidally from galaxies that were previously well separated, and dynamical modelling (Toomre \& Toomre \markcite{TT72}1972; Howard et al. \markcite{HKBB93}1993; Sofue \markcite{So94}1994; Elmegreen et al. \markcite{ESKBE95}1995a; Dubinski et al. \markcite{DMH96}1996; Mihos \& Hernquist \markcite{MH96}1996; Barnes \& Hernquist \markcite{BH96}1996; Mihos et al. \markcite{MDH98}1998; earlier work is reviewed by Barnes \& Hernquist \markcite{BH92a}1992) generally confirms that the tidal assumption is plausible although the simulations do not usually achieve the pronounced distension of the \ion{H}{1} relative to the stars that is often seen in the observed cases (however, see Smith et al. \markcite{SSP97}1997).

It has been suggested that some Ly$\alpha$ absorption lines, especially the Lyman Limit Systems (LLS), arise from tidally distended gas around individual galaxies (van Gorkom \markcite{vG93}1993) or within groups of galaxies (Morris \& van den Bergh \markcite{MvdB94}1994; Mo \& Morris \markcite{MM94}1994).
Such absorption systems account for a few percent of the low redshift Ly-$\alpha$ absorptions observed with HST (Storrie-Lombardi et al. \markcite{SMIH94}1994; Stengler-Larrea et al. \markcite{SBS+95}1995; Bahcall et al. \markcite{BBB+96}1996); whereas low redshift Ly-$\alpha$ forest systems are found in galaxy groups with a covering factor near unity (Le Brun et al. \markcite{lBBB96}1996; Bowen et al. \markcite{BBP96}1996; Morris et al. \markcite{MWD+93}1993), LLS are found much more rarely.
The \ion{H}{1} in interacting systems in any case serves as an historical record allowing the trajectories of the interacting galaxies to be traced back in time in some cases (Phookun et al. \markcite{PMTW92}1992; Chengalur et al. \markcite{CST95}1995; Elmegreen et al. \markcite{ESKBE95}1995).
Condensations of gas within the \ion{H}{1} tails appears in some instances to form new dwarf galaxies (Mirabel et al. \markcite{MDL92}1992; Elmegreen et al. \markcite{EKT93}1993; Duc \& Mirabel \markcite{DM94}1994; Hibbard et al. \markcite{HGGS94}1994; Hunsberger et al. \markcite{HCZ96}1996; Malphrus et al. \markcite{MSGH97}1997).

On the other hand, not all diffuse, extended \ion{H}{1} clouds appear to have been generated tidally from previously more compact galaxies.
Tidal effects may have shaped the Leo ring, but the progenitor does not appear to have been any of the optically identified galaxies in the group (Schneider \markcite{Sch89}1989).
The large, diffuse cloud \ion{H}{1}$\,$1225+01 (Giovanelli et al. \markcite{GWH91}1991) appears to be well isolated from any galaxy large enough to have exerted tidal influence.
In both cases the \ion{H}{1} gas appears to have never been part of any optically bright galaxy, and its dynamics appears to be controlled by dark matter potential wells of galaxy group scale rather than by tides due to individual galaxy halos (see also Hoffman et al. \markcite{HLSCF98}1998).

NGC$\,$4532 and  DDO$\,$137 were initially mapped by us at Arecibo\footnote{The Arecibo Observatory is part of the National Astronomy and Ionosphere Center, which is operated by Cornell University under a management agreement with the National Science Foundation.}
as two distinct irregular galaxies (Helou, Hoffman \& Salpeter \markcite{HHS84}1984; Hoffman et al. \markcite{HHSGS87}1987), but a chance reference beam for an unrelated observation lead to our discovery of a large, diffuse cloud of \ion{H}{1} surrounding the pair (Hoffman et al. \markcite{HSLR92}1992, \markcite{HLSFLR93}1993).
Presumably, if HST spectra were obtained for any QSOs lying behind the extended \ion{H}{1} cloud, anywhere within the outermost contours in our Arecibo map (see Fig. \ref{n4532prtothi}, below),  they would exhibit Lyman Limit Systems (LLS). 
Indications from the Arecibo maps of condensations within the diffuse cloud, but well outside the optical images of the two galaxies, motivated the Very Large Array\footnote{The Very Large Array of the National Radio Astronomy Observatory is a facility of the National Science Foundation, operated under cooperative agreement by Associated Universities, Inc.} maps presented here.
A desire for finer spatial resolution of the two main galaxies was another motivating factor.
CCD imaging was undertaken to see if any low surface brightness optical features could be found in the condensations within the diffuse cloud, and to investigate the star formation histories of the two galaxies.
As we shall detail below, we do not think that tides acting on the \ion{H}{1} reservoirs of two initially well-separated irregular galaxies can account for the diffuse \ion{H}{1} envelope.
This system seems to be another example of a primordial, star-poor cloud of \ion{H}{1} reacting to a clumpy group-scale dark matter halo.

Optical details culled mainly from de Vaucouleurs et al. \markcite{RC3}(1991, hereafter RC3) are presented in Table \ref{opttab}, which lists for each member galaxy the coordinates (epoch 1950) of the nucleus; the morphological type as coded by Binggeli et al. \markcite{BST85}(1985); the optical heliocentric velocity; our assumed distance, taken to be the same as the Virgo cluster core ; the blue apparent magnitude $B_{T}^{0}$ corrected for inclination and Milky Way extinction, from RC3; the blue absolute magnitude determined from that assumed distance and $B_{T}^{0}$; and the optical diameter at the 25 mag $[\arcsec]^{-2}$ isophote.

A brief description of the CCD imaging is given in Sect. \ref{CCD}.
Details and results of our \ion{H}{1} synthesis mapping are given in Sect. \ref{mapping}.
In Sect. \ref{discuss} we discuss the dynamics and possible evolutionary scenarios for the system.
We end with a summary and conclusions in Sect. \ref{sum}.

\section{\label{CCD}CCD imaging}

Both NGC$\,$4532 and DDO$\,$137 were observed in U, B, V, R and I 
with the Palomar 60-inch telescope equipped with a TK 2048x2048 CCD 
chip under photometric conditions.  For NGC$\,$4532, we also 
obtained an H$\alpha$ image.   These images will be fully described
in a later paper where we study the past and current star formation
activities in these galaxies (Lu, Hoffman, \& Salpeter \markcite{LHS98}1998, hereafter LHS98).  
Here, we make use of the B and R images only.   With a pixel size of 
$0.\arcsec742$, the CCD field is large enough for us to put two of 
the three HI  concentrations of the system (i.e., Clouds A and B; 
see Sect.~3.3) on the CCD image of DDO$\,$137, with Cloud C on the CCD frame for NGC$\,$4532.   The deep R image of DDO$\,$137 has an r.m.s. sensitivity of $\sim 26$ mag$\,$arcsec$^{-2}$.
But at this level, none of the HI clouds is detected optically.

Elliptical contours are fit to each of the galaxy images to extract 
a surface brightness profile, following the prescription given by Lu 
{\it et al.} (1993).  However, the inner parts of both galaxies 
are dominated by the light from a few clumps that are located off 
the galaxy center as determined by the isophotes at large radii.  As
a result, the fitted contours are dictated by these clumps at small
radii, causing the centers of the ellipses to shift. 
To avoid this problem, we adopted a hybrid method.  Namely,
the fitted result was used only at radii larger than $r_c$ where 
the surface brightness is smooth; but at radii less than $r_c$ 
the radial surface brightness profile was determined by using a series
of concentric elliptical contours with fixed center position, shape 
and position angle equal to the corresponding mean values of 
the fit between radii $r_1$ to $r_2$ in the outer galaxy.   Note that
the two latter parameters were also averaged over the B and R passbands.
In this ``concentric elliptical model'' method, the surface brightness
along an ellipse of radius $r_n$ is given by the mean value within 
the annulus between radii $0.5[r_n + r_{(n-1)}]$ and $0.5[r_n + 
r_{(n+1)}]$, where $r_{(n-1)}$, $r_n$ and $r_{(n+1)}$ are the radii 
of three successive ellipses.  The resulting hybrid surface brightness
profiles are shown in Fig.~\ref{n4532profs} for NGC$\,$4532 and DDO$\,$137.

For each galaxy, Table~\ref{CCDtab} gives the values of $r_c$, $r_1$ and $r_2$ as
well as a few parameters derived from the fit to the galaxy disk between
$r_1$ and $r_2$: the total magnitude in B, (B-R) color, the B-band 
isophotal diameters at B$\,$25 and B$\,$26$\,$mag$\,[\arcsec]^{-2}$, the mean ellipticity
and position angle averaged over the B and R passbands, the B-band 
exponential scale length $r_s$ as defined in $\mu(r) = \mu_0 + 
1.086 r/r_s$, and the B-band central surface brightness $\mu_0$.  
If appropriate, a roughly estimated r.m.s.~uncertainty is also given
in the table.

\section{\label{mapping}Neutral hydrogen observations}

The \ion{H}{1} mapping for these galaxies was conducted at the Very Large Array 
in two observing sessions as detailed in Table \ref{obsdet}, which lists for each source the date of the observing run, the pointing center (epoch 1950), the heliocentric velocity to which the receivers were tuned, the array configuration, the number of spectral channels and channel separation in velocity units, and the time spent on source during the observations, not including time spent on calibrators.
We have also listed the resulting beam size in arcsec, the rms noise per channel, and the $3\sigma$ detection limit for \ion{H}{1} column density in each channel.
We obtained both R and L circularly polarized data.
We found, however, that we could not eliminate striping of the C or D array maps as long as the R polarization data was retained, so we made maps using L polarization only.
On-line Hanning smoothing was employed throughout, and the observations were calibrated using sources from the VLA calibrator list.
Calibration and data-editing were accomplished using standard AIPS tasks.
Continuum was subtracted in the uv domain using the task UVBAS, and maps were made and CLEANed using IMAGR with zerospacing fluxes as discussed below and with robustness parameter 0 for the best compromise between spatial resolution and maximum signal-to-noise.
After imaging, each data cube was corrected for the VLA primary beam.

We estimated zerospacing fluxes from the Arecibo mapping by interpolating each spectrum between Arecibo velocity channels to find intensities at the velocities of the VLA channels.
Then we integrated over the entire spatial map at each interpolated velocity to produce the zerospacing flux estimate for each VLA channel.
The issue of what weight is appropriate for zerospacing fluxes in IMAGR does not appear to have been resolved yet, so we experimented with a wide range of weights.
We found that if the weight is made too large, the synthesized beam increases in size.
On the other hand, large weights would give us the best chance of recovering all the flux detected in the Arecibo map.
The weight we chose is the largest weight that does not increase the beamsize by more than
1\%, and the resulting spectrum integrated over the VLA map agrees well with the global spectrum integrated over the Arecibo results.

Following common practise, we have abbreviated the names so that the field ``N4532'' is centered on the optical image of NGC$\,$4532 while ``DDO$\,$137'' is centered a few arcmin west of the optical image of DDO$\,$137 in order to give maximum sensitivity to the \ion{H}{1} plume extending to the west of the galaxy as seen in our Arecibo mapping (Hoffman et al. \markcite{HSLR92}1992, \markcite{HLSFLR93}1993).
The field ``N4532/D137'' is centered in the region between the two galaxies to provide maximum sensitivity to \ion{H}{1} condensations within that region of the cloud enveloping the two galaxies.

We obtained data for two fields using the D array, one field centered on NGC$\,$4532 and the second nearly due S, a few arcmin W of the optical image of DDO$\,$137.
Since significant emission seems to be confined to the region of overlap in the two D array fields, we combined them as follows:
Each uv data set was first shifted to a common tangent point using AIPS task UVFIX; the sets were then combined using task DBCON.
IMAGR was run on the combined dataset, producing maps with a $46\arcsec \times 42\arcsec$ synthesized beam, 1.6 mJy ${\rm Bm}^{-1}$ rms, and a $3\sigma$ column density limit of $2.8 \times 10^{19}\;$atoms$\;{\rm cm}^{-2}$.
Finally, we applied a primary beam correction by averaging the standard VLA primary beam corrections for each original pointing center, giving each equal weight.
The result is a data cube that exhibits the same column densities near the centers of each galaxy, but which shows the diffuse emission more clearly than either of the separate fields.
We similarly combined the D array and C array datasets, weighting the primary beam corrections by the number of visibilities in each data set.
The result is a final set of maps with an $18.5\arcsec \times 16.7\arcsec$ synthesized beam and rms noise 0.58 mJy ${\rm Bm}^{-1}$ per channel, or a $3\sigma$ column density limit of $6.5 \times 10^{19}\;$atoms$\;{\rm cm}^{-2}$.
The C+D array mosaic is shown in Fig. \ref{n4532prCDpan}, where we have integrated over consecutive sets of three channels.
Earlier attempts to merge the datasets using the maximum entropy method task VTESS produced a large smear of apparently spurious emission across the center of the field; that is why we resorted to the method described above.
Efforts at merging both VLA datasets with our single-dish Arecibo map (Hoffman et al. \markcite{HLSFLR93}1993, hereafter HLS+93) are underway (Connell \markcite{Co98}1998), but our thrust here is to consider the VLA and Arecibo data in parallel.

In addition to emission from each of the two optical galaxies, we have clearly detected emission from three gas concentrations distinct from those galaxies (but still embedded within the low column density envelope we found in our Arecibo mapping [HLS+93]).
We will call these clouds A, B and C as labeled in Fig. \ref{n4532prtothi} where the 0th moment map for the entire region is superimposed on a Digitized Sky Survey\footnote{The Digitized Sky Surveys were produced at the Space Telescope Science Institute under U.S. Government grant NAG W-2166. 
The images of these surveys are based on photographic data obtained using the Oschin Schmidt Telescope, which is operated by the California Institute of Technology and Palomar Observatory on Palomar Mountain. 
The plates were processed into the present compressed digital form with the permission of that institution.} image.
Cloud B is part of the extended plume stretching W and S from DDO$\,$137 in the Arecibo map (HLS+93), which is evident as low level emission in each cube at velocities 1990--2052 km ${\rm s}^{-1}$.
Weak absorption against a central source in NGC$\,$4532 is evident at velocities $\leq 1979\;$km$\,{\rm s}^{-1}$; the feature is shown integrated over velocity in Fig. \ref{n4532abs}. 

\subsection{NGC$\,$4532}

The total \ion{H}{1} map, velocity field and second moment map of the SmIII galaxy NGC$\,$4532 are  shown in Figs. \ref{n4532tot}a, \ref{n4532vel}a and \ref{n4532disp}a.
The total \ion{H}{1} emission, integrated over the merged C+D array map, we find 50.9 Jy-km ${\rm s}^{-1}$, or $4.3 \times 10^{9} d_{19}^2\:{\rm M}_{\sun}$, where $d_{19}$ is the distance to the galaxy, assumed to be the same as that to the Virgo cluster, in units of 19 Mpc.
This is equal, within the uncertainties, to the flux we attribute to NGC$\,$4532 in our Arecibo mapping (see Table \ref{restab}; the fluxes reported in HLS+93 for NGC$\,$4532 / DDO$\,$137 did not properly correct for beam overlaps --- please note that this error affected only the map of NGC$\,$4532 / DDO$\,$137, not the other maps reported in HLS+93). 
The \ion{H}{1} distribution is displaced from the optical image of the galaxy in the direction of DDO$\,$137, and there is a flaring of \ion{H}{1} toward the E with a distinct tail of \ion{H}{1} extending from the S end of the galaxy toward the E.
The velocity field is similarly distorted; at the N end the contours are centered on the optical major axis and are consistent with a rising rotation curve symmetric about the optical center.
But to the S the contours are skewed, probably by tidal interaction, and the flaring \ion{H}{1} toward the E evidently is kinematically distinct from the rest of the \ion{H}{1} in the galaxy.

Although the observed velocity field does not constrain the position angle nor inclination of the innermost rings of NGC$\,$4532, we do obtain convergence, at values close to those determined optically, for the annulus having inner radius $50\arcsec$ and outer radius $70\arcsec$.
Consequently we adopt those values:  position angle of the receding end of the major axis $347\arcdeg$, inclination $72\arcdeg$, systemic velocity 2005 km ${\rm s}^{-1}$, for all rings of the galaxy.
The galaxy's center is not well determined by the optical data, due to an apparent very strong off-center starburst, and the outer portions of the \ion{H}{1} distribution and its velocity field are evidently perturbed by tidal interactions within the system.
The central part of the galaxy's velocity field should be least affected by external influences, so the systemic velocity was chosen to make the central part of the rotation curve symmetric.
The resulting rotation curve is displayed in Fig. \ref{n4532rotcur}a.
It appears to flatten on the approaching (SE) end, but continues to rise in more-or-less solid-body fashion on the receding end of the major axis.
The maximum ``rotation'' speed observed is 123 km ${\rm s}^{-1}$, on the receding side of the galaxy; the approaching side appears to level off at a velocity of about 58 km ${\rm s}^{-1}$.
If we average the rotation velocities within entire rings, as shown in Fig. \ref{n4532rcfit}a below, the maximum rotation speed is 102 km ${\rm s}^{-1}$ at a radius of $100\arcsec = 9.2$ kpc.
The second moment map indicates a rather uniform velocity dispersion within the C+D array beam, around 20 km ${\rm s}^{-1}$ over most of the map, decreasing to 10 km ${\rm s}^{-1}$ at the outskirts and with one large off-center region in which the dispersion rises above 30 km ${\rm s}^{-1}$.
Since our synthesized beam is large compared to the scale on which the rotation curve rises, we expect the central part of the map to overestimate the dispersion.
The dispersion is also likely to be overestimated around the outer edges of the map due to uneven blanking of the individual planes.
The relevant dispersion for dynamical considerations is that in the outer regions; we adopt a value of $\sigma_{z} = 24\;$km$\,{\rm s}^{-1}$, the average over the SE and NW portions of the second moment field avoiding the center and avoiding the outer edges.
This may still be a mild overestimate, but fortunately the dispersion is small compared to the rotation velocity so that errors in the dispersion have little effect on dynamical mass estimates.
The dynamical mass $M_{dyn} = ( v_{rot}^{2} + 3 \sigma_{z}^{2} ) r / {\rm G}$ is then $2.6 \times 10^{10} d_{19}\:{\rm M}_{\sun}$.

The surface brightness profiles of NGC$\,$4532 in B and R, from LHS98, are shown in Fig. \ref{n4532profs}a with the resulting stellar and \ion{H}{1} surface density profiles following in Fig. \ref{n4532surden}a. 
Neither the gas mass surface density (including the contribution from primordial helium by multiplying by a factor of $4/3$) nor the stellar mass surface density (obtained from the R surface brightness profile, assuming a constant ${M_*}/{L_R}$) is precisely exponential, but reasonable exponential fits can be made to each surface density distribution and to their sum.
In Fig. \ref{n4532rcfit}a, the lower dashed curve is the contribution to the rotation velocity from the gas, assuming it is in a thin disk.
As is evident in Fig. \ref{n4532tot}, the gas ``disk'' is far from thin and therefore produces a smaller rotation velocity for given \ion{H}{1} mass than the infinitessimally thin disk assumed in Fig. \ref{n4532rcfit}a.
Furthermore, the off-center starburst causes the center of the gas disk not to coincide with the apparent center of the stellar disk.
Instead of the usual ``maximum disk'' construction, we define a ``supermaximal stellar disk'' to give a rotation velocity curve close to the observed curve at intermediate radii (slightly exceeding the observed curve at radii $\leq 40\arcsec$ to allow for beam smearing), without any contribution from gas or dark matter.
This is intended to give the largest reasonable estimate of the stellar mass-to-light ratio, with the underestimate due to the thin disk assumption compensated by the overestimate due to the neglect of the gas.
The rotation velocity resulting from this ``supermaximal stellar disk'' is shown as the solid curve in Fig. \ref{n4532rcfit}a and requires a mass-to-light ratio ${M_*}/{L_R} = 0.24 {{\rm M}_{\sun}} / {{\rm L}_{\sun}}$ giving a total stellar mass $M_* = 2.4 \times 10^{9} d_{19}^2\:{\rm M}_{\sun}$.
This is a small ${M_*}/{L_R}$, but may be consistent with the indications of a strong starburst in the optical images (Bruzual \& Charlot \markcite{BC93}1993).
We will discuss this point further in LHS98.
If the rotation curve averaged within entire annuli of the galaxy is appropriate dynamically,  we need a significant contribution from dark matter in the outer parts of the galaxy.

\subsection{DDO$\,$137}

Corresponding maps of total \ion{H}{1} and velocity field for the SmIV galaxy DDO$\,$137, from the merged C+D array data 
cube, are displayed in Figs. \ref{n4532tot}b and \ref{n4532vel}b.
The integrated emission in the merged C+D map is 10.1 Jy-km ${\rm s}^{-1}$, or $8.6 \times 10^{8}  d_{19}^2\:{\rm M}_{\sun}$.
We cannot compare directly to the Arecibo map in this case, since the Arecibo mapping did not allow us to separate out DDO$\,$137 from the plume extending W and curving S from the galaxy. 
There are weak indications in the \ion{H}{1} map of a tidal streamer reaching toward the plume, but no strong distortions of the velocity field.
The relatively weak rotation of the galaxy appears approximately solid-body in form over most of the first-moment map.

The position angle from elliptical fits to the outer isophotes of DDO$\,$137 in LHS98 is $130\arcdeg$, but the kinematic major axis is apparently very close to EW.
Fitting the velocity field within consecutive annuli, with the inclination held fixed at $30\arcdeg$ since it is not well constrained by the \ion{H}{1} data, indicates that the position angle of consecutive rings gradually shifts from about $96\arcdeg$ for inner rings to about $82\arcdeg$ for the outermost couple.
To plot the rotation curve shown in Fig. \ref{n4532rotcur}b, we held inclination, center position and systemic velocity fixed for all rings, but allowed the position angle to be determined by the fit.
The rotation curve is probably best described as solid-body with perturbations.
The deviations from a solid-body curve should not be taken too seriously in model-fitting, however, since the \ion{H}{1} map indicates interaction effects.
In particular, the outermost rings may be sampling portions of the extended plume rather than DDO$\,$137 itself.
The second moment map gives an average velocity dispersion $\sigma_{z} = 15\;$km$\,{\rm s}^{-1}$ within the C+D array beam, avoiding the edges of the map which may be corrupted by blanking effects; this is presumably an overestimate since the beam is not small compared to the scale on which the rotation curve rises.
We will use $\sigma_{z} = 10\;$km$\,{\rm s}^{-1}$ for dynamical purposes.
If we can take at face value the highest measured rotation speed, $59 (0.5/\sin i )\;$km$\,{\rm s}^{-1}$, at $75.\arcsec2 = 6.9$ kpc on the receding side of the galaxy, farthest from the apparent tidal stream toward Cloud B, we find a dynamical mass $M_{dyn} =  5.6 [(0.5/\sin i)^{2} + 0.086 (\sigma_{z}/10)^{2}] \times 10^{9} d_{19}\:{\rm M}_{\sun}$.

The surface brightness profiles in B and R, from LHS98, and the \ion{H}{1} surface density profile of DDO$\,$137 are shown in Figs. \ref{n4532profs}b and \ref{n4532surden}b.
The gas mass surface density includes the contribution from primordial helium, by multiplying by a factor of $4/3$, and the stellar mass surface density is obtained from the R surface brightness profile, assuming a constant ${M_*}/{L_R}$. 
As for NGC$\,$4532, reasonable exponential fits can be made to each surface density distribution and to their sum.
In Fig. \ref{n4532rcfit}b we show the contributions of those exponential disks to the rotation curve of the galaxy, with ${M_*}/{L_R}$ adjusted to match the inner portion of the star-plus-gas disk to the corresponding portion of the rotation curve (exceeding the innermost two points to allow for beam-smearing and ignoring the thickness of the disks).
In this case ${M_*} / {L_R} = 2.2 {{\rm M}_{\sun}} / {{\rm L}_{\sun}}$ gives a reasonable fit to the entire rotation curve, except for the outermost point which is probably affected by non-circular gas motions.
There is perhaps no need for a dark matter halo in DDO$\,$137; however, the maximum disk value of ${M_*}/{L_R}$ is probably an overestimate if the considerations of Bottema (\markcite{Bo97}1997) outweigh the countereffect of the thickness of the gas and stellar disks.

\subsection{\label{clouds}Three gas concentrations outside the galaxies}

Cloud A, detected separately in the Arecibo spectra, the D array data and the C array data, is clearly visible in the total \ion{H}{1} map about $6\arcmin$ E of NGC$\,$4532, $7\arcmin$ due N of DDO$\,$137.
The total \ion{H}{1} emission from the cloud, summed over the channels (1906-1958$\;$km$\;{\rm s}^{-1}$) that show line emission at that location, is shown in Fig. \ref{cloudtot}a.
It is resolved by our C+D array data, but is only a few beamwidths across.
There is some indication of rotation (or tidal shear) in the velocity field, with the NW end having lower velocity than the SE end, but the end to end difference is only about two of our velocity channels and the contours are far from regular in shape.

Cloud B is a concentration within the extended plume that extends W, then curves S, from DDO$\,$137 in our Arecibo map (HLS+93).
It is visible as a distinct cloud in the D array and C array data separately, in the velocity range 1906-1990$\;$km$\;{\rm s}^{-1}$.
The total \ion{H}{1} map is shown in Fig. \ref{cloudtot}b; it is clearly resolved by our C+D array data.
The velocity field does not display any regular pattern, although there are larger velocity gradients within Cloud B than within Cloud A.

Cloud C (Figs. \ref{cloudtot}c) appears at lower velocity than any other feature in the maps, in the range 1823-1927$\;$km$\;{\rm s}^{-1}$ channels.
It is visible in the Arecibo data as well as in the C and D array data separately.
As with the other two clouds, there is no clear order to its velocity field.

\section{\label{discuss}Discussion}

\subsection{Comparisons to Arecibo spectra}

Table \ref{restab} compares measured quantities from the VLA mapping reported here to corresponding quantities from our Arecibo mapping reported in HLS+93.
Unfortunately, the integrated \ion{H}{1} fluxes and hydrogen masses given in Table 2 of HLS+93 for the components of the NGC$\,$4532 / DDO$\,$137 system did not correctly account for beam overlaps; the values cited in Table \ref{restab} of this paper supercede those reported earlier.
(This error in HLS+93 affected only NGC$\,$4532 / DDO$\,$137, not the other galaxies discussed there.)
With the heavily-weighted zerospacing fluxes used here,  the VLA integrated \ion{H}{1} fluxes  for NGC$\,$4532 agrees well with the \ion{H}{1} emission integrated over the Arecibo spectra attributed to that galaxy.
For DDO$\,$137, the Arecibo spatial resolution was not adequate to separate the galaxy from the extended plume which contains Cloud B, so direct comparison of total fluxes is not meaningful.

The integrated spectra from the merged D+C array data cube also agree quite well in detail with the single beam Arecibo spectra acquired when Arecibo was pointed at the optical centers of the two galaxies.
Fig. \ref{n4532d137spec} compares the spectra for each of the galaxies.
Clearly measurements of systemic velocity and profile widths made from these spectra would agree within a fraction of the VLA channel spacing as shown in Table \ref{restab}.
For NGC$\,$4532, half the Arecibo profile width at 50\% of the nearest peak (80.5 km ${\rm s}^{-1}$) is a bit smaller than the maximum rotation velocity inferred from the VLA map ($102\:\sin 72\arcdeg = 97\;$km$\,{\rm s}^{-1}$).
The inferred systemic velocities do not agree quite so well:  2020 km ${\rm s}^{-1}$ from the midpoint of the two 50\% points of the Arecibo profile compared to 2005 km ${\rm s}^{-1}$ from fitting the velocity field in the VLA map.
For DDO$\,$137, half the Arecibo profile width at 50\% of peak is 33.4 km ${\rm s}^{-1}$ while the maximum rotation velocity from the VLA map gives $59\:\sin 30\arcdeg = 30\;$km$\,{\rm s}^{-1}$.
The systemic velocity taken to be the midpoint of the points where the Arecibo profile falls to 50\% of its peak is 2066 km ${\rm s}^{-1}$, slightly larger than the value reported for DDO$\,$137 plus plume in Table \ref{restab}; we inferred a value of 2057 km ${\rm s}^{-1}$ from the VLA velocity field.

Integrated spectra for each of the three gas concentrations are compared to the Arecibo spectra from the closest pointings in Fig. \ref{cloudspec}.
The integration in each case extends over the largest contiguous contour in Figs. \ref{cloudtot}a, b, or c.
Here the agreement is not so good, because the Arecibo beam is not pointed at the center of the cloud in each case and has collected flux from other parts of the plume or from the outskirts of the galaxies.
In the case of Cloud A, the bump in the Arecibo spectrum extending up to 2100 km ${\rm s}^{-1}$ is apparently due to incompletely subtracted sidelobes of NGC$\,$4532.
For Cloud B, confusion with DDO 137 and the extended plume makes it impossible for us to separate out an \ion{H}{1} mass for the cloud alone from Arecibo spectra, but for Clouds A and C we can interpolate among the Arecibo spectrum nearest to each cloud's position and the six nearest neighbor beam positions (with some judgment by eye of what portion of each spectrum to attribute to the cloud).
The resulting estimated \ion{H}{1} masses, with uncertainties $> 20$\%, are displayed in Table \ref{restab}.

\subsection{Indicative masses of the three gas concentrations}

The extreme velocities encountered within Cloud A are 1927 and $1952\;$km$\,{\rm s}^{-1}$, at opposite ends of the cloud separated by about $72\arcsec = 6.6 d_{19}\;$kpc.
This suggests rotation, but the isovelocity contours are far from regular.
The second moment map has an average velocity dispersion $\sigma_{z} = 7.5\;$km$\,{\rm s}^{-1}$, an overestimate since there is a velocity gradient on a scale comparable to the beamsize.
If we could assume that the cloud is in simple rotation and that our perspective is edge-on --- neither assumption is likely to be true --- we would have a dynamical mass $M_{dyn} = ( v_{rot}^{2} + 3 \sigma_{z}^{2} ) r / {\rm G} = 2.5 \times 10^{8} d_{19}\:{\rm M}_{\sun}$.
However, the assumption of rotation is suspect; tidal shear is equally likely to be the cause of the velocity gradient.
Consequently we report in Table \ref{dynmas} only the indicative gravitational mass $M_{ind} = 3 ({\Delta V_{50}}/2)^{2} r / G = 6.2 \times 10^{8} d_{19}\:{\rm M}_{\sun}$ (which would be the dynamical mass if $\Delta V_{50}/2$ were a good measure of the line-of-sight component of the turbulent velocity within the cloud and if the cloud were in Virial equilibrium).
The integrated \ion{H}{1} amounts to 1.5 Jy$\;$km$\,{\rm s}^{-1}$, or $1.3 \times 10^{8} d_{19}^{2}\:{\rm M}_{\sun}$, so the \ion{H}{1} mass alone is probably not sufficient to bind the cloud if the Virial theorem estimate approximates its true mass.

For Cloud B, which we take to be the largest distinct clump near the center of Fig. \ref{cloudtot}b, the situation is even less clear.
The extreme velocities are approximately 1980 and $2050\;$km$\,{\rm s}^{-1}$, but the low velocity occurs in the center of the cloud.
It would be incorrect to associate this velocity field with circular rotation.
The width of the integrated spectrum, Fig. \ref{cloudspec}b, is $91\;$km$\,{\rm s}^{-1}$ at the points 50\% down from the peaks of the double-horned profile.
The velocity dispersion, from the average over the second moment map in regions where the velocity field is quiet, is $\sigma_{z} = 19\;$km$\,{\rm s}^{-1}$, but that overestimates the local gas velocity dispersion as discussed above.
The linear dimension of the cloud is $113\arcsec$, or $10.4 d_{19}\;$kpc.
The indicative gravitational mass is then $M_{ind} =  3 ({\Delta V_{50}}/2)^{2} r / {\rm G} = 7.5 \times 10^{9} d_{19}\:{\rm M}_{\sun}$.
The integrated \ion{H}{1} mass is 4.5 Jy-km ${\rm s}^{-1}$, or $3.9 \times 10^{8} d_{19}^{2}\:{\rm M}_{\sun}$.
If Cloud B is gravitationally bound, evidently much dark matter is required.
But the assumption that it is in Virial equilibrium is highly suspect, since the cloud lies in an extended plume.

Cloud C likewise does not have a well-ordered velocity field.
Within the cloud itself, velocities range from about 1854 to 1894 km$\;{\rm s}^{-1}$, but in this case it is the high velocities which occur near the center of the cloud.
The profile width at 50\% of the peak of the spectrum, Fig. \ref{cloudspec}c, is $76\;$km$\,{\rm s}^{-1}$.
The beam-scale line-of-sight dispersion $\sigma_{z}$ is about 17 km ${\rm s}^{-1}$, averaged over the second moment map in regions where the gradient in the first moment is small.
The linear size of the cloud is $126\arcsec = 11.6 d_{19}$ kpc.
Taking half the profile width to indicate turbulent velocity as for Clouds A and B, above, we obtain an estimate of the indicative gravitational mass $M_{ind} =  3 ({\Delta V_{50}}/2)^{2} r / {\rm G} = 5.8 \times 10^{9} d_{19}\:{\rm M}_{\sun}$.
The integrated \ion{H}{1} mass is 3.1 Jy-km ${\rm s}^{-1}$, or $2.6 \times 10^{8} d_{19}^{2}\:{\rm M}_{\sun}$.
The dark matter requirements are about as extreme as for Cloud B, if Cloud C is to be bound, but again the assumption of Virial equilibrium is suspect since Cloud C also lies within an extended stream.

The \ion{H}{1} masses and dynamical masses estimated above for all three clouds are very much consistent with those quantities for the faintest Im galaxies in Virgo for which we have \ion{H}{1} detections (Hoffman et al. \markcite{HHSGS87}1987).

\subsection{Interaction scenarios and group mass estimates}

At a cursory glance, neither the \ion{H}{1} distribution of NGC$\,$4532 taken in isolation nor that of DDO$\,$137, each shown in Fig. \ref{n4532tot}, alerts the viewer to the complex interactions within the system.
Only upon much closer inspection does one notice the asymmetry of the \ion{H}{1} around NGC$\,$4532, with a more rapid decrease in column density toward the N than to the S; the flare to the E does not reveal itself until one inspects the individual channel maps (Fig. \ref{n4532prCDpan}) or the velocity field (Fig. \ref{n4532vel}a).
Another possible explanation for the asymmetrical \ion{H}{1} distribution around NGC$\,$4532 might be ram pressure by the Virgo cluster intracluster medium
(Phookun \& Mundy \markcite{PM95}1995).
However, the galaxy is not close to any X-ray emitting portion of the cluster complex, nor to any particularly dense subcluster.
Furthermore, the absorption seen against a continuum source toward the N end of the optical image is a clue that some unusual starburst is going on (LHS98).
It seems more likely that the disturbance is caused somehow by interactions within the NGC$\,$4532 / DDO$\,$137 system. 
Ram pressure against diffuse \ion{H}{1} within the system is a possibility, but the well-separated gas concentrations and strange kinematics of the diffuse gas lead us to the more likely possibility that a clumpy dark matter halo enveloping the entire system is driving the motions and concentration of the gas.
Neither the column density map (Fig. \ref{n4532tot}b) nor the velocity field (Fig. \ref{n4532vel}b) of DDO$\,$137 is unusual for an irregular galaxy; it is only in the larger context of a map showing DDO$\,$137 and Cloud B along with the plume curving to the SW that one realizes that the slight linear extension toward the NW of DDO$\,$137 may in fact be a stream.
There is, however, no hint of a stellar tidal stream emanating from either galaxy in the B- or R-band CCD images from LHS98, shown in Fig.~\ref{n4532tot}b.
What might be taken for a tidal tail stretching SE from NGC$\,$4532 in the DSS image of the POSS-I plate is revealed to be only a string of individual foreground stars on the POSS-II film and in our CCD frames.

Could another galaxy be involved?
There is no other catalogued galaxy with known redshift $< 3000\;$km$\,{\rm s}^{-1}$ within $34\arcmin$.
The nearest large galaxy is NGC$\,$4570, an ${\rm S}0_{1}(7)/{\rm E}7$, $B_{T} = 11.84$, galaxy at $62\arcmin$ (more than 5 times the separation of DDO$\,$137 from NGC$\,$4532) and $V_{\sun} = 1730\;$km$\,{\rm s}^{-1}$ (RC3).
On the Digitized Sky Survey it appears to be a classic edge-on S0 with no hint of tidal interaction.

Could some portion of the extended gas distribution be produced by a blowout from the intense starburst in NGC$\,$4532 (e.g. Meurer et al. \markcite{MSK98}1998)?
A blowout toward us along the line-of-sight would account for the blue-shifted velocities of Clouds A and C.
However, gas is not expected to achieve escape velocity from a galaxy as massive as this (Mac Low \& Ferrara \markcite{MLF98}1998), and the blowout would more likely occur perpendicular to the disk of the galaxy which we see nearly edge-on.
The gas would also have to have time to collect and condense into the observed clouds, which is unlikely to happen while the starburst is still underway.
In any case, we would be left with the question of what provided the trigger for the starburst.

We see no hint of starlight at our R sensitivity limit of $26\;$mag$\,[\arcsec]^{-2}$ at the positions of the clouds in any of the other CCD images (LHS98); the B-band images are shown in Figs. \ref{cloudtot}a, b, and c, above.
(We take the off-center smudge in Cloud A to be a background galaxy; it is indistinguishable from the many other background galaxy ``smudges'' throughout the frame.)
Are these gas concentrations nascent dwarf galaxies, eventually to form stars as in the tidal streams studied by Mirabel et al. \markcite{MDL92}(1992), Elmegreen et al. \markcite{EKT93}(1993), Duc \& Mirabel \markcite{DM94}(1994), Hibbard et al. \markcite{HGGS94}(1994), Hunsberger et al. \markcite{HCZ96}(1996), and Malphrus et al. \markcite{MSGH97}(1997)?
The \ion{H}{1} masses reported in the preceding section are appropriate to observed dwarf star-forming galaxies (Salpeter \& Hoffman \markcite{SH96}1996; Roberts \& Haynes \markcite{RH94}1994), but the \ion{H}{1} mass alone does not appear to be sufficient to gravitationally bind any of the three clouds if they are currently in Virial equilibrium.
Dwarf galaxies formed from tidal debris are not expected to have significant dark matter halos (Barnes \& Hernquist \markcite{BH92b}1992b; Elmegreen et al. \markcite{EKT93}1993).
The central column densities are $5-6\times 10^{20}\;$atoms$\;{\rm cm}^{-2}$ for each, close to the critical column density for star formation (Kennicutt \markcite{Ke89}1989), but our surface brightness limits are close to the central surface brightnesses of the most extreme low surface brightness (LSB) galaxies known (O'Neil et al. \markcite{OBC97}1997), and more than $2\;{\rm mag}[\arcsec]^{-2}$ fainter than most LSB galaxies.

We can estimate the mass of the group of 5 objects (2 galaxies and 3 clouds) on the assumption that the system is in Virial equilibrium either by taking each of the five as a point object tracer of the gravitational potential of the group, using a standard Virial estimator (e.g., Binney \& Tremaine \markcite{BT87}1987) or by measuring the velocity width of the global \ion{H}{1} profile.
Both approaches give consistent results:  a group mass in the range $3-6 \times 10^{11}\:{\rm M}_{\sun}$ (see Table \ref{dynmas}).
This is approximately an order of magnitude larger than the sum of the dynamical masses of all the components, estimated from their rotation curves or \ion{H}{1} profile widths.
Table \ref{dynmas} also reports the total stellar mass of each luminous component, calculated from the $R$ magnitude with RC3 corrections for internal and Galactic extinction, using the disk $M / L_{R}$ from our rotation curve fitting.
The gas mass of each distinct component is also tabulated; it is calculated from the \ion{H}{1} mass (VLA measurement for each galaxy or cloud, Arecibo measurement for the system as a whole) by multiplying by a factor of $4/3$ to account for primordial Helium gas.
We have ignored any contribution from molecular gas.
Unless the Virial estimates of the group mass are deceiving us by more than a factor of a few, the dynamics of the group are controlled by dark matter which lies mostly outside the objects we can see.

On a similar vein, the total \ion{H}{1} mass of the group as measured at Arecibo is larger by 60\% than the sum of the \ion{H}{1} masses of the two galaxies as measured by the VLA --- in other words, the extragalactic gas amounts to 75\% of the gas remaining in NGC$\,$4532, or more than 3 times the gas remaining in DDO$\,$137; and 28\% of the total \ion{H}{1} lies outside all of the galaxies and discrete gas concentrations.
Furthermore, as we noted in HLS+93, the bulk of the extragalactic gas --- Clouds A and C along with the southwestern part of the plume stretching out from DDO$\,$137 and the diffuse material extending both E and W from NGC$\,$4532, visible in the Arecibo map but only hinted at in the VLA maps presented here --- falls at lower velocity than either of the two galaxies.
Only Cloud B lies at the velocity of NGC$\,$4532, below that of DDO$\,$137 to which it is apparently connected.
Yet these features are spread out in all directions around the two galaxies.
Any one of these features appearing at lower velocity than its associated galaxy would not be surprising; a tidal stream which achieved escape velocity mainly along the line-of-sight would appear blueshifted.
But it would be highly improbable for the aggregate --- one plume, two discrete clouds and a large quantity of diffuse gas --- {\em all} to be blueshifted if they were produced by tides acting between the two galaxies.
Nor is it likely that any tidal scenario can account for such a large fraction of the gas having been removed from the galaxies.

Sancisi (\markcite{Sa98}1998) has suggested that, when a disk galaxy shows signs of recent interaction but has no companion galaxy of sufficient size visible today, a companion may have been present before the interaction but has now dispersed (or has been incorporated into the larger disk).
Hunter et al. (\markcite{HWvWGK98}1998) make a similar suggestion for NGC$\,$4449.
Because of the large quantity of very extended gas present in our system, we are instead led to speculate that the supposed tidal features are the result of the interaction between one large Sm galaxy (NGC$\,$4532) and a large \ion{H}{1} cloud resembling \ion{H}{1}$\,$1225+01 (Giovanelli et al. \markcite{GWH91}1991), each with a large complement of dark matter (or perhaps in a common halo of dark matter).
The dark matter haloes are now overlapping, with the more diffuse \ion{H}{1} cloud having been disrupted in the process; star formation has proceeded only in the most dense clump (DDO$\,$137) of the fragmented cloud.
A star formation history analysis of CCD imaging in UBVRI and H$\alpha$  is underway (LHS98).
Clouds A, B and C may either be manifestations of the tidal interaction between the original \ion{H}{1} cloud and the clumpy dark matter potential, or the result of gas falling into dense clumps of dark matter within the common halo.
High resolution simulations show that such clumps can survive for significant times within group-scale halos, if their internal densities are sufficiently high (Klypin et al. \markcite{KGKK97}1997; Tormen et al. \markcite{TBW97}1997a, \markcite{TDS97}b; Ghigna et al. \markcite{GMG+98}1998).

\section{\label{sum}Summary}

Neutral hydrogen synthesis array mapping of the extended gas cloud around the widely separated galaxy pair NGC$\,$4532 / DDO$\,$137 reveals three \ion{H}{1} concentrations well separated from the two main galaxies.
The \ion{H}{1} masses and indicative gravitational masses are consistent with those of faint Im galaxies in the Virgo cluster (Hoffman et al. \markcite{HHSGS87}1987).
No stars are visible within these clouds on the CCD images of Lu \markcite{LHS98}(1998).
The inferred dynamical mass of each cloud, assuming Virial equilibrium, exceeds the \ion{H}{1} mass by factors of 4-20.
The central column density of each cloud is close to the critical column density for star formation (Kennicutt \markcite{Ke89}1989), but our upper limits on surface brightness are comparable to the central surface brightnesses of the most extreme low surface brightness  galaxies known (O'Neil et al. \markcite{OBC97}1997).

The \ion{H}{1} distribution and rotation curve of NGC$\,$4532 display some peculiarities consistent with tidal interaction; those of DDO$\,$137 are more nearly in accord with expectations for a low mass irregular galaxy.
A maximum disk hypothesis applied to fitting exponential star+gas disks to the observed rotation curves would require very different mass-to-light ratios for the stars in NGC$\,$4532 and those in DDO$\,$137, perhaps consistent with their very different apparent star formation activities (Lu et al. \markcite{LHS98}1998).
The shapes of the rotation curves of DDO$\,$137 and of the approaching side of NGC$\,$4532 are consistent with a constant ratio of dark matter to star+gas density at all observed radii in each; however, if the receding side of NGC$\,$4532 is the one less severely afflicted with non-circular velocities (as suggested by all other indications of tidal interaction in the galaxy), then the usual dark matter halo is required.

All available estimators of the total mass of the group (consisting of two galaxies plus three extragalactic clouds) give a total mass an order of magnitude larger than the sum of the masses of the individual objects out to the largest radii measured in the VLA maps, as estimated from their rotation curves or Virial analyses.
The \ion{H}{1} fluxes reported in Hoffman et al. (\markcite{HLSFLR93}1993) for NGC$\,$4532 and DDO$\,$137 were not properly corrected for beam overlaps; we give the corrected fluxes in Table \ref{restab}.
(The error affected only NGC$\,$4532 / DDO$\,$137, not the other galaxies discussed in that earlier paper.)
After correction, the total flux at Arecibo still exceeds the sum of the \ion{H}{1} masses of the two galaxies as measured in the VLA mapping by 37\%, and it exceeds the sum of \ion{H}{1} masses of all discrete components (including Clouds A, B, and C) by 28\%.
These facts, along with the kinematic peculiarities of the system, lead us to advance the hypothesis that the system progenitor was a bound pair consisting of the large Sm galaxy NGC$\,$4532 and a large star-poor gas cloud similar to \ion{H}{1}$\,$1225+01, each dominated by dark matter.
The dark matter halos have now merged, and gas is being accreted into the densest clumps within the common dark matter halo.
Star formation has proceeded only in the two densest clumps --- NGC$\,$4532, which probably had stars prior to the merger, and DDO$\,$137, which perhaps had no prior stars.
Detailed star formation history analysis to follow (Lu et al. \markcite{LHS98}1998) should help us to constrain this scenario further.

\acknowledgments

We thank J.H. Loh and A. Rabbani for assistance with some preliminary data reductions and the staff of the Very Large Array for their assistance in acquiring the \ion{H}{1} synthesis data.
The anonymous referee's voluminous comments lead to substantial improvements in our analysis and presentation.
This work was supported in part by US National Science Foundation grants AST-9015181 and AST-9316213 at Lafayette College.

\begin{table}
\dummytable\label{opttab}
\end{table}

\begin{figure}
\plotone{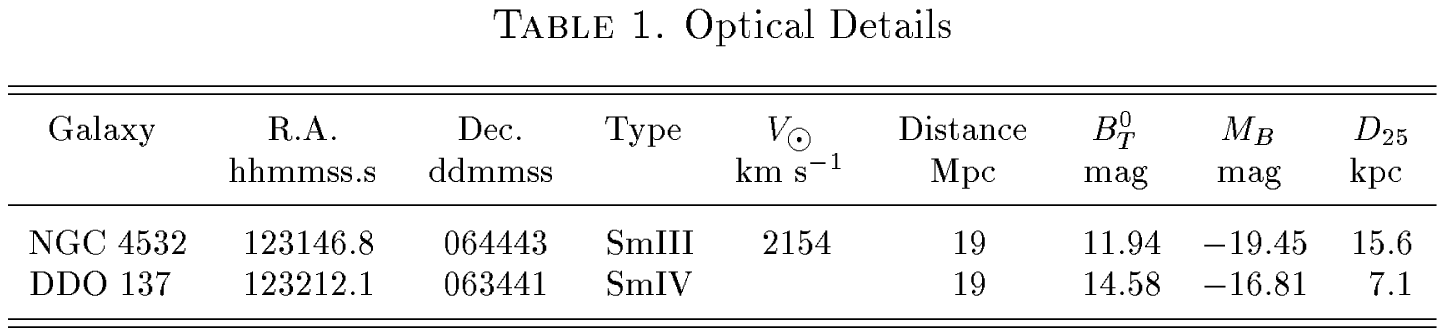}
\end{figure}

\begin{table}
\dummytable\label{CCDtab}
\end{table}

\begin{figure}
\plotone{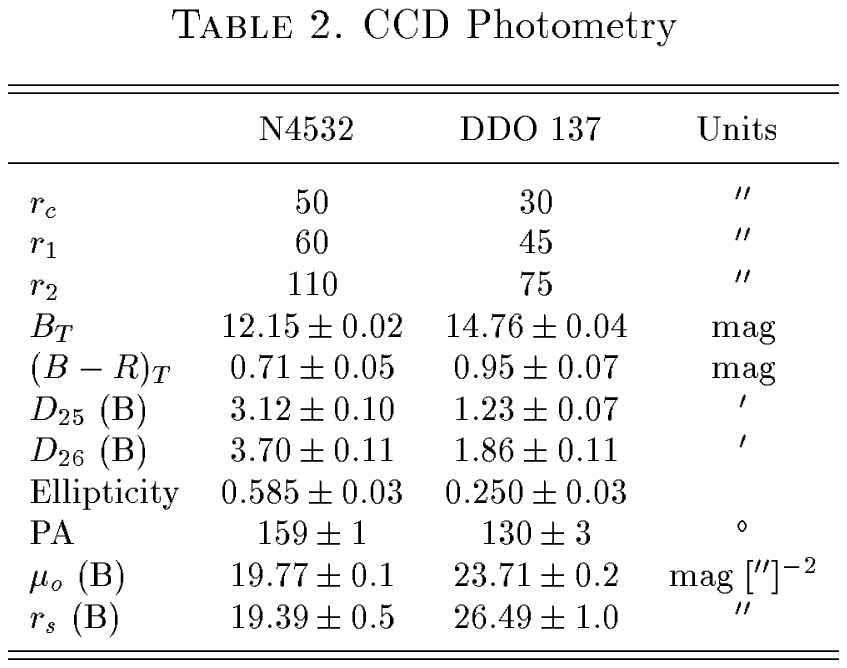}
\end{figure}

\begin{table}
\dummytable\label{obsdet}
\end{table}

\begin{figure}
\plotone{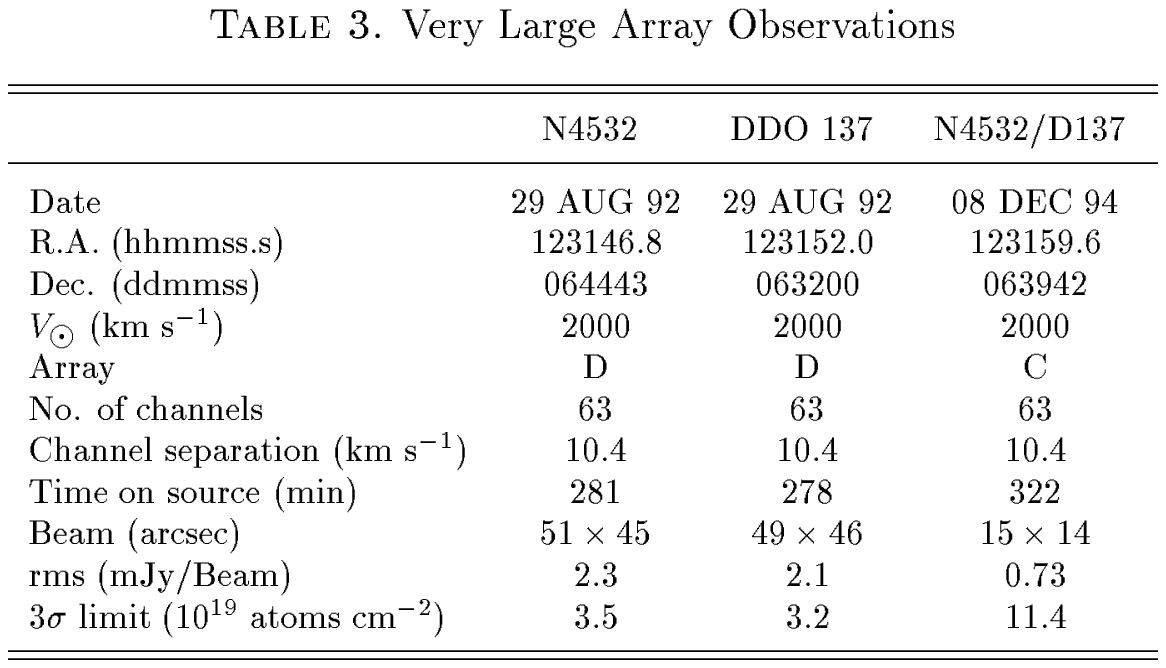}
\end{figure}

\begin{table}
\dummytable\label{restab}
\end{table}

\begin{figure}
\plotone{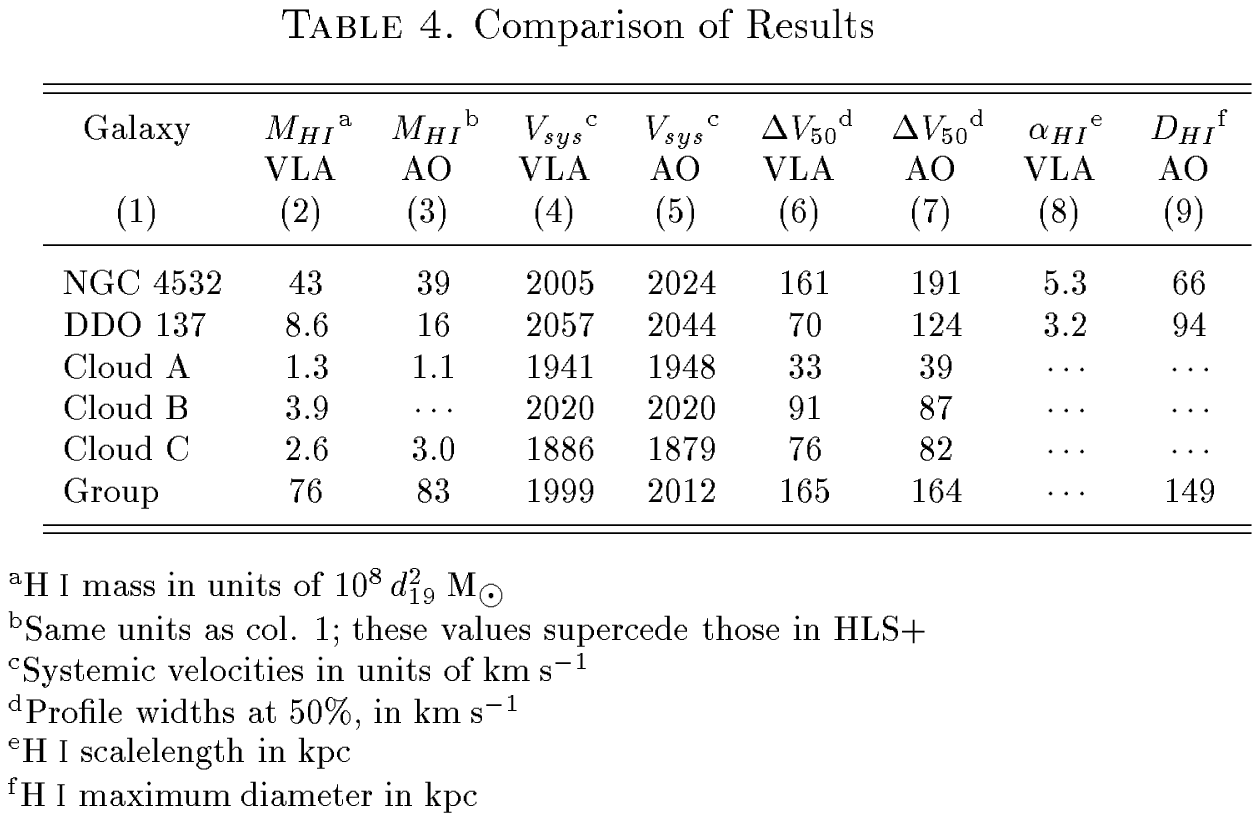}
\end{figure}

\begin{table}
\dummytable\label{dynmas}
\end{table}

\begin{figure}
\plotone{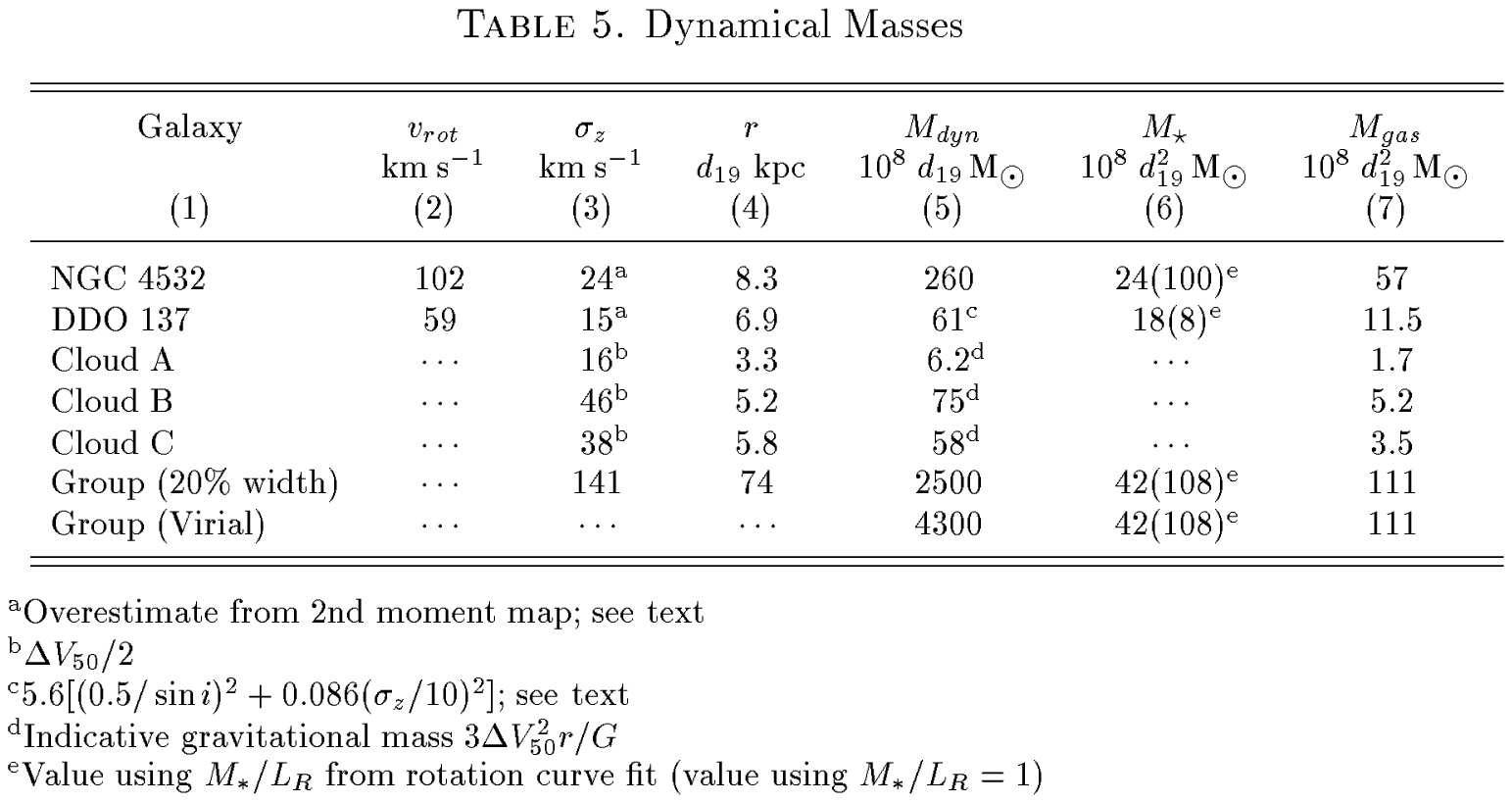}
\end{figure}

\clearpage

\begin{figure}
\caption{
\label{n4532prCDpan}
Mosaic of 3-channel integrated contour maps for the merged C and D array datasets of the galaxy pair NGC$\,$4532 / DDO$\,$137.
The beam is shown in the top left corner of the top left pane.
Each pane consists of the integral over three channels of the data cube; the velocities of the middle channel in each pane, in the order in which they are numbered, are 2146, 2115, 2084, 2052, 2021, 1990, 1958, 1927, 1896, 1864, 1833 and 1802 km ${\rm s}^{-1}$.
Contours are drawn at $-3.6$, $-1.8$, 1.8, 3.6, 7.2 and 14.3$\times 10^{20}\;$atoms$\;{\rm cm}^{-2}$.}
\end{figure}

\begin{figure}
\caption{
\label{n4532prtothi}
Contours of total hydrogen emission from the area around NGC$\,$4532 and DDO$\,$137, integrated over all line-bearing channels for 
the C+D array combination and superimposed on a greyscale image from DSS.
Contour levels from the VLA mapping ({\em solid line}) are 
1.8, 3.6, 7.2, 14.3, 28.6 and 57.2$\times 10^{20}\;$atoms$\;{\rm cm}^{-2}$.
The dashed curves show the outermost contour levels from the Arecibo map, at 2 and 4$\times 10^{19}\;$atoms$\;{\rm cm}^{-2}$.
Three prominent concentrations of gas outside the main galaxies are indicated.}
\end{figure}

\begin{figure}
\caption{
\label{n4532abs}
Absorption feature at the center of NGC$\,$4532, integrated over velocity for the channels in which it appears in the combined C+D array data cube.
The emission in those channels (excluding all channels where no absorption is seen) is shown with solid contours, absorption with dashed contours.
The contour levels are $-0.40$, $-0.20$, $-0.10$, 0.10, 0.20 and 0.40 Jy ${\rm Bm}^{-1}\;$km$\,{\rm s}^{-1}$.}
\end{figure}

\begin{figure}
\caption{
\label{n4532tot}
Contours of total hydrogen emission from NGC$\,$4532 ({\em left}) and DDO$\,$137 ({\em right}), integrated over all line-bearing channels from  our combined C+D cube superimposed on R-band CCD images from Lu et al. (1998).
Neutral hydrogen contour levels are $-7.2$, 7.2, 14.3, 28.6 and 57.2$\times 10^{20}\;$atoms$\;{\rm cm}^{-2}$ for NGC$\,$4532 and $-3.6$, 3.6, 7.2 and 14.3$\times 10^{20}\;$atoms$\;{\rm cm}^{-2}$ for DDO$\,$137.
The CCD images employ a logarithmic transfer function, with the stretch chosen to show faint outer features without completely saturating the central parts of the galaxy.}
\end{figure}

\begin{figure}
\caption{
\label{n4532vel}
Isovelocity contours superimposed on a greyscale image of the velocity field from the merged C+D data of NGC$\,$4532 ({\em left} panel) and DDO$\,$137 ({\em right} panel).
Contours are drawn from 1955 to 2090 in steps of $15\;$km$\,{\rm s}^{-1}$ for NGC$\,$4532, and from 2030 to 2090 in steps of $10\;$km$\,{\rm s}^{-1}$ for DDO$\,$137.}
\end{figure}

\begin{figure}
\caption{
\label{n4532disp}
Second moment contours superimposed on a greyscale image of the second moment velocity field from the merged C+D data  of NGC$\,$4532 ({\em left} panel) and DDO$\,$137 ({\em right} panel).
Contours are drawn from  10 to 40 in steps of $10\;$km$\,{\rm s}^{-1}$ for NGC$\,$4532 and from 5 to 20 in steps of $5\;$km$\,{\rm s}^{-1}$ for DDO$\,$137.}
\end{figure}

\begin{figure}
\caption{
\label{n4532rotcur}
Rotation curves for NGC$\,$4532 ({\em left} panel) and DDO$\,$137 ({\em right} panel), determined by fitting the velocity fields within consecutive annuli.
For NGC$\,$4532, the center position, systemic velocity (2005 km$\;{\rm s}^{-1}$), position angle and inclination were held fixed for all rings at values discussed in the text.
Positive velocities are receding and positive radii are toward the NW.
For DDO$\,$137, the center position, systemic velocity (2057 km$\;{\rm s}^{-1}$), and inclination were held fixed for all rings at values discussed in the text; the position angle was determined by the fit.
Positive velocities are receding and positive radii are toward the E.
The cross in each panel marks the center in position and rotation velocity.}
\end{figure}

\begin{figure}
\caption{
\label{n4532profs}
B ({\em triangles}) and R ({\em circles}) surface brightness profiles for NGC$\,$4532 ({\em left panel}) and for DDO$\,$137 ({\em right panel}) from Palomar CCD imaging.}
\end{figure}

\begin{figure}
\caption{
\label{n4532surden}
Mass surface densities in stars ({\em triangles}) and gas ({\em circles}) for NGC$\,$4532 ({\em left panel}) and DDO$\,$137 ({\em right panel}).
We have assumed ${M_*} / {L_R} = 1 {{\rm M}_{\sun}} / {{\rm L}_{\sun}}$ for the purposes of these plots, and have included neutral hydrogen and primordial helium gas but have neglected molecular and ionized gas.}
\end{figure}

\begin{figure}
\caption{
\label{n4532rcfit}
The rotation curves ({\em points}) of NGC$\,$4532 ({\em left panel}) and DDO$\,$137 ({\em right panel}) averaged over entire rings.
The light dashed curve in each panel shows the contribution of a thin exponential disk fitted to the gas, including neutral hydrogen and primordial helium.
For NGC$\,$4532 the solid curve shows the contribution of the ``supermaximal'' thin exponential disk fitted to the stars assuming ${M_*} / {L_R} = 0.24$ (see text).
For DDO$\,$137 the light dotted curve shows the contribution of a thin stellar exponential disk having  ${M_*} / {L_R} = 2.2 {{\rm M}_{\sun}} / {{\rm L}_{\sun}}$ while the solid curve shows that of a thin exponential disk fit to the sum of stars plus gas.}
\end{figure}

\begin{figure}
\caption{
\label{cloudtot}(Figs. 11a-c are only available via WWW at http://www.lafayette.edu/faculty/hoffmang.)
Contours of total hydrogen emission from Clouds A, B, and C, integrated over all line-bearing channels from our C+D array combination (1906-1958$\;$km$\;{\rm s}^{-1}$ for A, 1906-1990$\;$km$\;{\rm s}^{-1}$ for B, and 1823-1927$\;$km$\;{\rm s}^{-1}$ for C).
The contours are overlaid on greyscale images of B-band CCD frames from Lu (1997).
Contour levels are $-2.1$, 2.1, 3.2 and 4.8$\times 10^{20}\;$atoms$\;{\rm cm}^{-2}$ for A, $-3.6$, 3.6, 5.4, 8.0 and 12.1$\times 10^{20}\;$atoms$\;{\rm cm}^{-2}$ for B, and $-2.9$, 2.9, 4.3 and 6.4$\times 10^{20}\;$atoms$\;{\rm cm}^{-2}$ for C.
The stretch in each logarithmic B-band image is chosen to bring out the faintest possible features.
Traces of DDO$\,$137 and NGC$\,$4532 remain near the left-hand edges of the frames for B and C, respectively.
The triangular ``paper-clip'' artifact just S of NGC$\,$4532 in the image for Cloud C is spurious.}
\end{figure}

\begin{figure}
\caption{
\label{n4532d137spec}
Spectra integrated over the individual galaxies ({\em open circles}) compared to the Arecibo central beam profile ({\em solid line}) for NGC$\,$4532 ({\em left}) and DDO$\,$137 ({\em right}).}
\end{figure}

\begin{figure}
\caption{
\label{cloudspec}
Spectra integrated spearately over Clouds A, B, and C  ({\em open circles}) compared to the Arecibo single beam profile ({\em solid line}) from the beam position closest to each cloud.}
\end{figure}

\begin{figure}
\caption{
\label{globalspec}
Global spectrum of the NGC$\,$4532 / DDO$\,$137 system integrated over the Arecibo map ({\em solid curve}) with VLA contributions from each of the discrete components indicated:  NGC$\,$4532 ({\em filled circles}), DDO$\,$137 ({\em filled squares}), and Clouds A, B, and C ({\em open triangles, squares and pentagons respectively}).
The dashed curve is the sum of the VLA spectra for the five discrete components.}
\end{figure}

\end{document}